\documentclass[epj,nopacs]{svjour}
\usepackage{amsmath,amssymb,amsfonts,graphicx,cite,multirow,color,algorithmic,algorithm}
\usepackage[utf8]{inputenc}
\usepackage{amsmath}
\usepackage{caption}
\usepackage{subcaption}
\usepackage{hyperref}
\graphicspath{{graphics/}}
\makeatletter
\ifx\input@path\@undefined
\def\input@path{{graphics/}}
\else
\g@addto@macro\input@path{{graphics/}}
\fi
\makeatother

\usepackage{snapshot}

\preprint{UWTHPH-2021-16\\MCnet-21-13\\LU-TP 21-45}

\title{Soft QCD Effects in VBS/VBF Topologies}

\author{Carsten Bittrich\inst{1}, Patrick Kirchgae\ss er\inst{2},
  Andreas Papaefstathiou\inst{3}, Simon Pl\"atzer\inst{4,5,6} and
  Stefanie Todt\inst{1}}
\authorrunning{C. Bittrich et al.}

\institute{Institut für Kern- und Teilchenphysik, TU Dresden, Dresden,
  Germany \and Theoretical Particle Physics, Department of Astronomy
  and Theoretical Physics, Lund University, Lund, Sweden \and
  Department of Physics, Kennesaw State University, Kennesaw, GA
  30144, USA \and Institute of Physics, NAWI Graz, University of Graz,
  Universit\"atsplatz 5, A-8010 Graz, Austria \and Particle Physics,
  Faculty of Physics, University of Vienna, Boltzmanngasse 5, A-1090
  Wien, Austria \and Erwin Schr\"odinger Institute for Mathematics and
  Physics, University of Vienna, Boltzmanngasse 9, A-1090 Wien,
  Austria}

\date{\today}

\abstract{We consider the impact of multi-parton interactions, colour
  reconnection and hadronization on the modeling of vector boson
  fusion and vector boson scattering (VBS) final states at the Large
  Hadron Collider (LHC). We investigate how the variation of the model
  parameters, compatible with a reasonable spread of predictions around
  typical tuning observables, extrapolates into the VBS phase
  space. We study the implications of this variation on the total uncertainty
  budget attached to realistic simulation of the final states in current
  event generator predictions. We find that the variations have a
  non-trivial phase space dependence and become comparable in size to
  the perturbative uncertainties once next-to-leading order
  predictions are combined with parton shower evolution.}

\begin{document}

\maketitle

%% start contents %%%%%%%%%%%%%%%%%%%%%%%%%%%%%%%%%%%%%%%%%%%%%%%%%%%%%%%%%%%%%%
%%%%%%%%%%%%%%%%%%%%%%%%%%%%%%%%%%%%%%%%%%%%%%%%%%%%%%%%%%%%%%%%%%%%%%%%%%%%%%%%

\section{Introduction}
\label{sec:Introduction}

An accurate description of the complex final states at the LHC is
vital to distinguish New Physics from those already encompassed by the
Standard Model. After the discovery of the Higgs boson, one focus of
the ongoing and future runs of the LHC is to precisely unravel the
physics of electroweak Higgs boson production, and the interactions of
weak gauge bosons themselves at high energies. This is key to
understanding the mechanism of electroweak symmetry breaking and the Higgs
boson properties, beyond those established so far.

While the processes of interest, vector boson fusion (VBF) into a
Higgs boson, or vector boson scattering (VBS) more generally, are
genuine electroweak processes, at a hadron collider, they will be
triggered at significant rates only if the strongly-interacting quarks
are involved, and as such QCD dynamics is unavoidable within this
class of processes. In particular, the expected final state consists
of a pair of highly-energetic jets originating from the quarks that
initiated the electroweak process, and little hadronic activity is
expected elsewhere due to the lack of exchange of colour charge in
between the so-called tagging
jets~\cite{Rainwater:1999sd,Rauch:2016pai}. These VBF, or VBS
signatures (we will not distinguish between VBF and VBS in the
following), however, will necessarily interfere with QCD-induced
processes, and even as fixed-order electroweak processes, they will
not necessarily be dominated by contributions from diagrams with a VBS
topology, {\it i.e.} a colour-less exchange in the $t$-channel of the
hard scattering. They will also interfere with processes that have
initial state ($s$-channel) or $u$-channel colour connections, as
well. These effects can be shown to be suppressed kinematically in the
phase space region exhibiting the VBS signatures by including a veto
on additional hadronic activity in between the tagging jets
\cite{Campanario:2018ppz}.

Accepting the VBS approximation, {\it i.e.} neglecting the
interferences and other topologies within the acceptance of interest,
as a reliable way of predicting the final states, one indeed finds
stunningly small QCD uncertainties at fixed order
\cite{Oleari:2003tc,Jager:2006zc}, even including a third jet {\it
  e.g.} in VBF Higgs production \cite{Figy:2007kv}. A similar statement
is true when including the effects of jet evolution as commonly
implemented in multi-purpose event generators and matched to
next-to-leading order (NLO) QCD predictions
\cite{Ballestrero:2018anz}. The jet radius dependence of the VBS cross
sections has also been studied carefully in a quest for more
perturbative stability. Numerically significant effects of QCD colour
singlet exchanges not covered by NNLO calculations in the $t$-channel
approximation, nor included in readily available parton shower
algorithms\footnote{Exceptions of this kind are newly developed
evolution algorithms at the amplitude level, see {\it e.g.}
\cite{DeAngelis:2020rvq}.} have been pointed out in the
literature~\cite{Campanario:2018ppz}. However, for the currently-available state of the art, one would consider these processes to be
theoretically accurate and precisely modeled at NLO QCD, at least
within the typical region of VBS acceptance. The merging of different
jet multiplicities has recently been addressed in VBF Higgs
production~\cite{hoeche2021study,Chen:2021phj}, and detailed studies of
parton shower effects are available in these
cases~\cite{Jager:2020hkz,Buckley:2021gfw}, suggesting that the
perturbative component is mostly well under control,
though~\cite{Chen:2021phj} has in fact pointed out that the role of
the VBF approximation in shower initial conditions for multi-jet
merging might be more delicate than naively expected.

In this work, we point out that gaining control of the perturbative
part of the prediction and the realistic simulation of the final
states of interest is a necessary ingredient to an uncertainty budget
of theoretical modeling, however by far not a sufficient one. We
therefore address the effect of dynamics in a proton-proton collision
which has not been accounted for in studies performed so far and are
typically outside a ready-to-use concept of estimating the impact of
neglected contributions and missing higher order corrections: those
are the physics of multi-parton interactions, hadronization, and
colour reconnection which we believe represent specifically important
effects, particularly in light of the distinct character of the VBS
processes, as well as the open question of the role of non-trivial QCD
effects originating from colour
evolution~\cite{Forshaw:2007vb,AngelesMartinez:2018cfz}.

While one would in fact expect that the overall impact of the physics
addressed in this work would not lead to tremendous effects, care
needs to be taken that variations within these contributions might
become comparable to the level of uncertainty reached by the
perturbative description, including a non-trivial phase space
dependence. This is something we need to accept will happen even in
light of minimum bias and underlying event data of small uncertainties
predicted very well by these models. It is the purpose of the present
work to explicitly check for the size of variations of the
non-perturbative models and to clarify whether their impact relative
to the perturbative part sets a reliable scale on the uncertainty one
should expect.

This manuscript is structured as follows: In
Sec.~\ref{sec:Unvertainties} we will review the overall uncertainty
budget of an event generator simulation, mostly following the lines
of~\cite{Bellm:2016rhh}, however with a focus on multi-parton
interaction (MPI) modeling, and the role of phenomenological models of
colour reconnection and hadronization. In Sec.~\ref{sec:Simulation} we
will then outline the simulation used in this study, which has been
mainly based on the VBFNLO
library~\cite{Baglio:2014uba,Baglio:2011juf,Arnold:2008rz} and the
Herwig 7 event generator~\cite{Bahr:2008pv,Bellm:2015jjp}, though we
do not expect that any of our initial reasoning and findings are
indeed specific to the models implemented within Herwig and would
equally well apply to any other event generator. In Sec.~\ref{sec:MPI}
we study the impact of MPI, and variations of the MPI model, onto VBS
final states for the first time, before we present conclusions and an
outlook into future work.

\section{Overview on Event Generator Uncertainties}
\label{sec:Unvertainties}

While there are very well established methods of estimating the
uncertainty of fixed-order and analytically resummed perturbative
calculations due to missing higher orders, the same concept does not
exist in one-to-one correspondence for even the perturbative parts of
event generators. Progress has been made, though, in varying the
scales involved in the parton shower evolution as well as the matching
algorithms \cite{Cormier:2018tog,Bellm:2016rhh} and can now be deemed a
reliable measure of parton shower uncertainty, included along with
variations in the hard process. A strong analytic statement is not
possible in these cases, owing to the fact that analytic insight into
parton shower-evolved observables is typically lacking.

What parton shower variations typically assess is the reliability of
predictions in phase space regions populated by parton shower
emissions, and how a poor description of such regions is improved by
matching and merging. To this extent, we can speak of reliable
perturbative input. However, as the perturbative parton shower
evolution does bridge the gap to the small scales at which
phenomenological models take over, these variations cannot make up the
entire uncertainty budget of an event generator and need to be
confronted with the reliability of models of soft physics --
describing multi-parton interactions, colour reconnection and
hadronization -- and their interplay needs to be carefully
evaluated. On the perturbative side, to be precise, we build on the
findings of~\cite{Rauch:2016jxo}, and use LO- and NLO-matched
simulations, varying the shower hard scale but not the profile
function by which the hard end of shower emissions' phase space is
approached, and using the `resummation profile' advocated in
\cite{Bellm:2016rhh} to preserve the properties we demand by a
combination of hard process and parton shower. We do not consider any
parametric uncertainties such as variations in the strong coupling or
parton distribution functions.

In the remainder of this section we will focus on MPI and colour
reconnection models, which are vital to describe the complexity
observed in hadronic collisions. The parameters of these models, to
be discussed in detail below, are typically tuned to observables which
are designed to probe the additional activity introduced by
multi-parton interactions, at scales smaller than the typical momentum
transfer of the hard scattering of interest. By virtue of their
abundant occurrence, these observables have very small experimental
uncertainty, and models deliver enough parameters and dynamics 
to deliver a decent fit of those data. The goodness-of-fit can be
quantified using standard methods and is now conveniently available in
form of so-called `Eigentunes'~\cite{Buckley:2009bj}. However, those
measures of uncertainty cannot reflect how precise we believe these
models are, nor how accurate they are outside the range of
observables typically considered for the soft physics tunes.

The MPI model is crucial for ``dressing'' the signal process with
additional hadronic activity. The underlying event activity is modeled
as perturbative QCD $2 \rightarrow 2$ processes and additional soft
interactions simulated as multiperipheral particle production
\cite{Bahr:2008wk,Bahr:2008dy,Bahr:2009ek,Gieseke:2016fpz,Bellm:2019icn}.
Here we give a quick overview of the main concepts of
interest for the discussion in later sections.  At the beginning of a
run with Herwig, the MPI model determines a matrix containing the
probabilities for the different number of $h$ hard and $n$ soft
interactions
\begin{align}
    P_{h,n}(s) = \frac{\sigma_{h,n}(s)}{\sigma_{\mathrm{inel}}(s)} \;,
\end{align}
where $\sigma_{\mathrm{inel}}(s)$ is the inelastic non-diffractive cross section
and $\sigma_{h,n}(s)$ is the cross section for $h$ hard and $n$ soft 
events
\begin{align}
    \sigma_{h,n}(s) =& \int d^2\textbf{b} \,\frac{\langle n(b,s) \rangle^h_{\mathrm{hard}}}{h!}
                                            \frac{\langle n(b,s)\rangle^n_{\mathrm{soft}} }{n!}\nonumber \\
                                        &\times e^{-(\langle n(b,s) \rangle_{\mathrm{hard}}  
                                        + \langle n(b,s) \rangle_{\mathrm{soft}} )}\;.
\end{align}
The number of soft and hard interactions follow a Poissonian distribution with mean values
\begin{align}
    \langle n(b,s) \rangle_{\mathrm{hard}} = A(b,\mu_{\mathrm{hard}})\,\sigma_{\mathrm{hard}}^{\mathrm{inc}}(p_{\perp}^{\mathrm{min}},s)
\end{align}
and 
\begin{align}
    \langle n(b,s) \rangle_{\mathrm{soft}} = A(b,\mu_{\mathrm{soft}})\,\sigma_{\mathrm{soft}}^{\mathrm{inc}}(s)\;.
\end{align}
$A(b,\mu)$ is the overlap function given by
\begin{align}
    A(b,\mu) = \frac{\mu^2}{96\pi} (\mu b)^3 K_3(\mu b)\;, 
\end{align}
where $K_3$ is the modified Bessel function of the second
kind. The overlap function has a different form for soft and hard
interactions since we assume that the interactions depend on different
matter distributions inside the proton.
$\sigma_{\mathrm{hard}}^{\mathrm{inc}}(p_{\perp}^{\mathrm{min}},s)$ is
the inclusive cross section for QCD $2 \rightarrow 2$ processes above
$p_{\perp}^{\mathrm{min}}$, calculable in perturbative QCD.  The soft
inclusive cross section $\sigma_{\mathrm{soft}}^{\mathrm{inc}}(s)$ and
the soft inverse proton radius $\mu_{\mathrm{soft}}$ are chosen such
that the total cross section $\sigma_{\mathrm{tot}}(s)$ is correctly
described.\footnote{The total cross section is determined with the
Donnachie-Landshoff parametrization \cite{Donnachie:1992ny}.}

The Herwig MPI model has two main genuine free parameters. 
These are the hard inverse proton radius $\mu_{\mathrm{hard}}$ and the 
minimum transverse momentum $p_{\perp}^{\mathrm{min}}$, which factorizes 
hard and soft interactions in terms of $p_{\perp}$. 

For the accurate description of minimum bias data and flavour
observables additional non-perturbative effects like colour
reconnection must be taken into account~\cite{Gieseke:2012ft,
  Gieseke:2017clv}.  Colour reconnection is used in this regard to
restore the notion of a pre-confined state~\cite{Gieseke:2012ft} in
the presence of extreme event topologies where many overlapping
clusters from multiple parton interactions are encountered.  After the
evolution of the parton shower has terminated, colour-preconfinement
states that colour connected partons are close in momentum space,
which leads to a distribution of invariant cluster masses peaking
at small values of $M$.  The \textit{plain} colour reconnection
algorithm in Herwig tries to find configurations of clusters that
reduce the sum of invariant cluster masses
\begin{equation}
    \lambda = \sum_{i=1}^{N_{cl}}M_i^2.
\end{equation}
The algorithm picks a cluster randomly from the list of clusters and
compares it to all other clusters in the event. The invariant mass of
the alternative cluster configuration is calculated and, if this leads
to a reduction in the sum of cluster masses, this configuration is
accepted with a fixed probability $p_{\mathrm{Reco}}$.

Models represent our lack of exact knowledge about the physics in
question in the non-perturbative regime.  There is no clear true/false
distinction between models possible. Rather, the evaluation of a
model's quality follows along the lines of better/worse.  A
puzzling question in the context of these models, which are often
inspired by theoretical principles concerns the correct assessment of
uncertainties.  The assessment of uncertainties for non-perturbative
models is in general an ill-defined problem.

\begin{figure}[t]
\centering
\includegraphics[width=8cm]{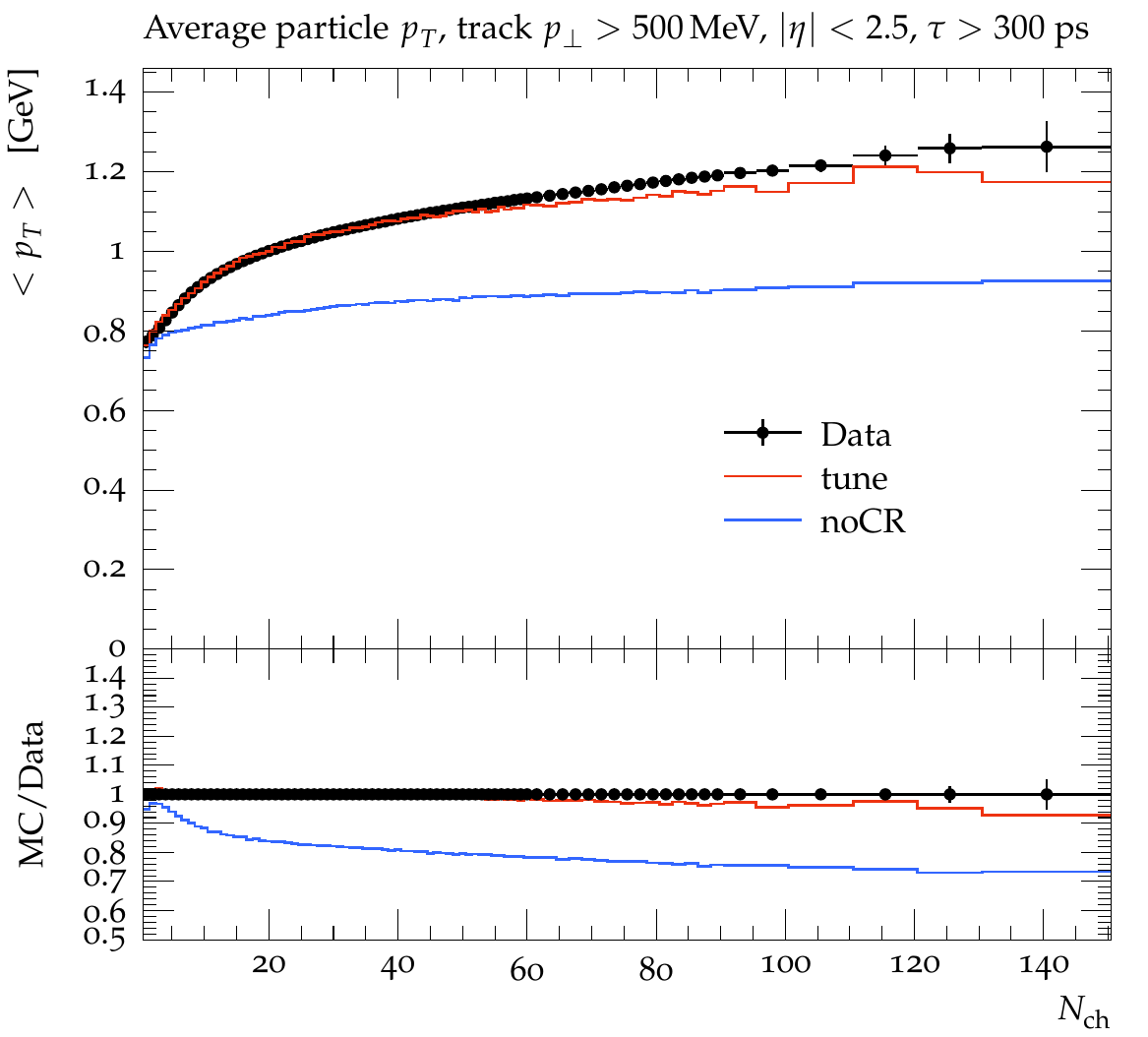}
    \caption{$ \langle p_{\perp} \rangle$ vs. $N_{\mathrm{ch}}$. Measured by \cite{Aad:2016mok}. 
    The distribution is not described without a colour reconnection mechanism.
  }
  \label{fig:CRoff}
\end{figure}

``On-off'' studies that involve the relevant parts of the model are frequently
used, for example to assess the impact of colour reconnection
or any other model that modifies the final state of the
simulation\cite{Juste:2013dsa}.  While this is a sensible approach to
study the effects of the model, minimum bias data clearly shows that
colour reconnection is a necessary ingredient in order to describe the
data (see fig.~\ref{fig:CRoff} and \cite{Argyropoulos:2014zoa}).
Therefore, turning off colour reconnection means ignoring important
physics effects, which implies that the on-off technique is
not suited to discuss the propagation of model uncertainties to VBF
observables.

\section{Outline of the Simulation}
\label{sec:Simulation}

As an example process we consider Vector Boson Fusion $Z$ production,
{\it i.e.} $pp\to \mu^+\mu^- + 2 \,{\text{jets}}$, since this is one
of the most relevant processes within current analyses and allows us
to closely follow the experimental acceptances. In this work we
consider the `signal' process mediated by electroweak interactions at
tree level and supplemented by NLO QCD corrections in the VBF
approximation as implemented in the VBFNLO library. VBFNLO has been
interfaced to Herwig 7 and we perform the matching to NLO QCD using
the Matchbox framework \cite{Platzer:2011bc} through the MC@NLO-type
matching paradigm to the angular ordered shower. In what follows, we
will refer to these simulations as `NLO'. As a cross check on the
simulation of the perturbative structures, we also consider a
simulation using tree-level merging based on MadGraph-generated
amplitudes~\cite{Alwall:2014hca} and merged with the dipole shower of
Herwig following the method outlined in
\cite{Platzer:2012bs,Bellm:2017ktr}. This simulation has the same --
leading order -- accuracy for the description of the third jet, and we
can further test if the VBF approximation is valid in this study. It
is vital that we can reduce the uncertainties of the modeling
of the third jet to actually observe the impact of soft QCD
variations on jet veto observables. These would otherwise be
mismodeled as reflected in large parton shower variations in the case
of the third jet not predicted by an exact matrix element.

Since we are considering the VBF approximation for most of this
study, we do require to set up the analysis using the VBF phase space
region of two energetic tagging jets at large invariant mass and
rapidity separation. Since this selection is not enforced as strictly as
one would feel comfortable with theoretically, we chose a tight and a
loose VBF cut setup, to probe the transition region towards
which the VBF approximation would not be considered to be very
reliable. To be specific, both sets of cuts require jets to be defined
with the anti-$k_\perp$ algorithm \cite{Cacciari:2008gp}, with
\begin{equation}\label{eq:jetcuts}
  p_{\perp,j} > 25~{\rm GeV}\;,\qquad |y_j| < 4.5\;,
\end{equation}
and a minimum of two jets. For these two jets, the `loose'
setup requires
\begin{equation}\label{eq:loosecuts}
  m_{12} > 250~{\rm GeV}\;,\qquad \Delta y_{12} > 0.0 \ ,
\end{equation}
while the `tight' setup requires
\begin{equation}\label{eq:tightcuts}
  m_{12} > 1000~{\rm GeV}\;,\qquad \Delta y_{12} > 2.0 \  .
\end{equation}
In order to evaluate the perturbative uncertainties, we vary the
renormalization and factorization scales by a factor $\sqrt{2}$ around their
central value, which are given by the $H_\perp$ of the hard scattering
\begin{equation}
  \mu_{R,F} = M_{\perp}(Z) + \sum_{i \in \text{jets}} p_{\perp,i} \ .
\end{equation}
These scales, but not their variations, set the hard scale above which
shower emissions are vetoed for the shower not to produce radiation
above the typical scales of the jets involved in the hard process. The
transition towards the hard phase space is implemented using the
`resummation profile' as discussed in \cite{Bellm:2016rhh}, and the
shower hard scale is then varied individually with $\pm
\sqrt{2}\mu_{R,F}$ to obtain an estimate on how reliable the shower
predictions are in certain regions.

\begin{figure}[t]
\centering
\includegraphics[width=8cm]{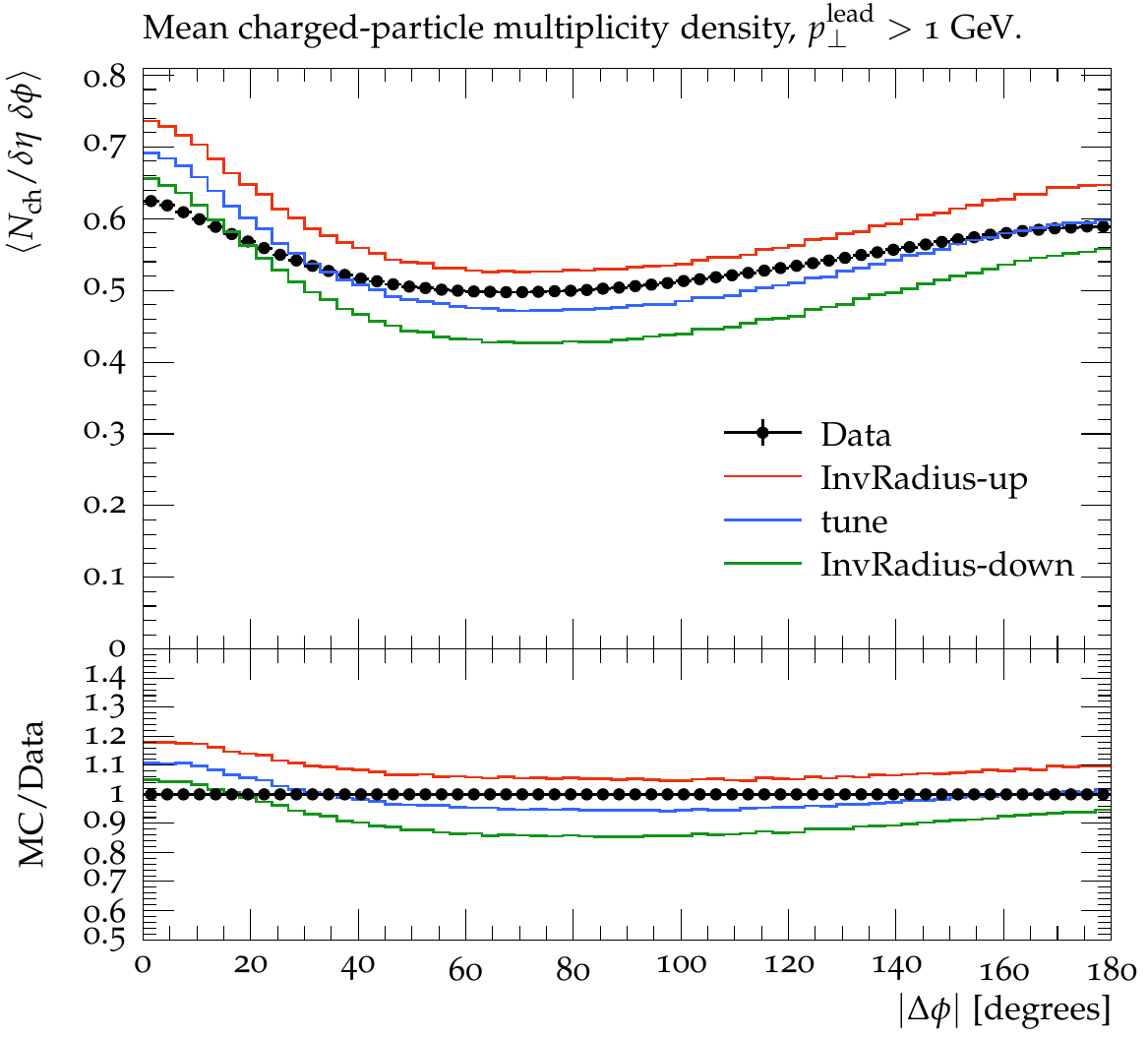}
\caption{ Plot of the mean charged particle multiplicity with respect
    to the difference in the azimuth angle $\Delta\Phi$ \cite{ATLAS:2017blj} for the tuned
    parameter values and the $\mu_{\mathrm{hard}}$ variations. The
  variations cover roughly a $\sim$ 10\% band around the description
  with the tuned parameter values.  }
\label{fig:UEvariation}
\end{figure}

As far as the non-perturbative variations are concerned, we follow the
approach to tune the MPI and colour reconnection model to 13 TeV
underlying event data from Ref. \cite{ATLAS:2017blj} and then find
variations in the parameter space which correspond to a $\sim$ 10$\%$
band around the tuned values.  We then proceed to study how these
parameter variations propagate to perturbative VBF observables. The tuned
parameter values and the corresponding parameter variations are shown
in Tab. \ref{tab:variations}.
\begin{table}
\centering
\begin{tabular}{c c c c } 
      \hline
      & tune & up & down   \\ [0.5ex] 
 \hline
     $p_{\perp}^{\mathrm{min}}$ & 3.19   & 3.4 & 2.9  \\ 
     $\mu_{\mathrm{hard}}$      & 0.92   & 1.2 & 0.75 \\ 
     $p_{\mathrm{Reco}}$        & 0.63   & 0.8 & 0.4  \\ 

 \hline

\end{tabular}
\caption{Tuned parameter values and parameter variations.}
\label{tab:variations}
\end{table}
By varying the parameters we still have all the physical mechanisms
incorporated in the simulation. We performed LO an NLO simulations for
the full set of parameter variations. The results of these simulations,
including additional observables to those presented in this article, as
well as the \texttt{Rivet} analysis~\cite{Buckley:2010ar} code used
can be found at the accompanying web page~\cite{cernpage}.
We find that the up and down variations of the inverse proton radius
$\mu_{\mathrm{hard}}$ constitutes the outermost variations of the third-leading jet VBF observables. 
Therefore, this parameter is especially suited for our
purposes since it is directly correlated with the amount of
underlying event activity induced from MPI processes (see 
fig.~\ref{fig:UEvariation}).

\section{Impact of Multi-parton Interactions}
\label{sec:MPI}

We have performed an investigation of the net impact of the different model components
on the final result of the simulation. In particular, we found that the
contribution of multi-parton interactions is significant and rises
with increasing jet radius, as one would expect. This is especially
true for distributions of the third jet, and possibly higher jet
multiplicities, which we have not been investigating in great
detail. We observe this both at leading, as well as next-to-leading
order matched simulation which implies that we do not mix model
contributions with a lack of higher order corrections.
In the following subsections we show the uncertainties in VBF
observables due to the outermost variations in the MPI model, which
corresponds to the variation of the inverse proton radius
$\mu_{\rm{hard}}$ as explained in Sec. \ref{sec:Simulation}.

\subsection{LO versus NLO}

\begin{figure*}[ht]
  \centering
  \includegraphics[width=0.48\textwidth]{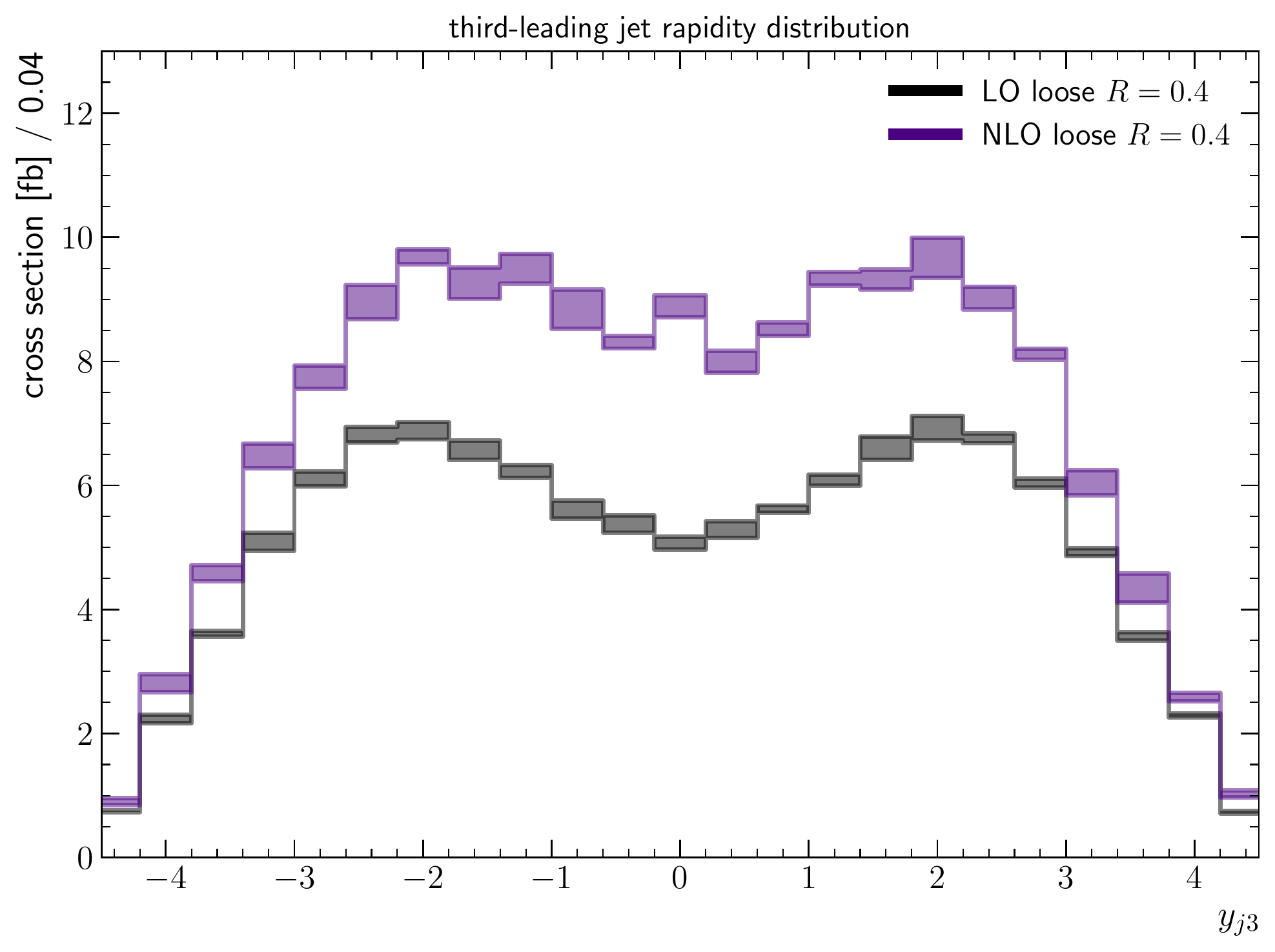}\hfill
  \includegraphics[width=0.48\textwidth]{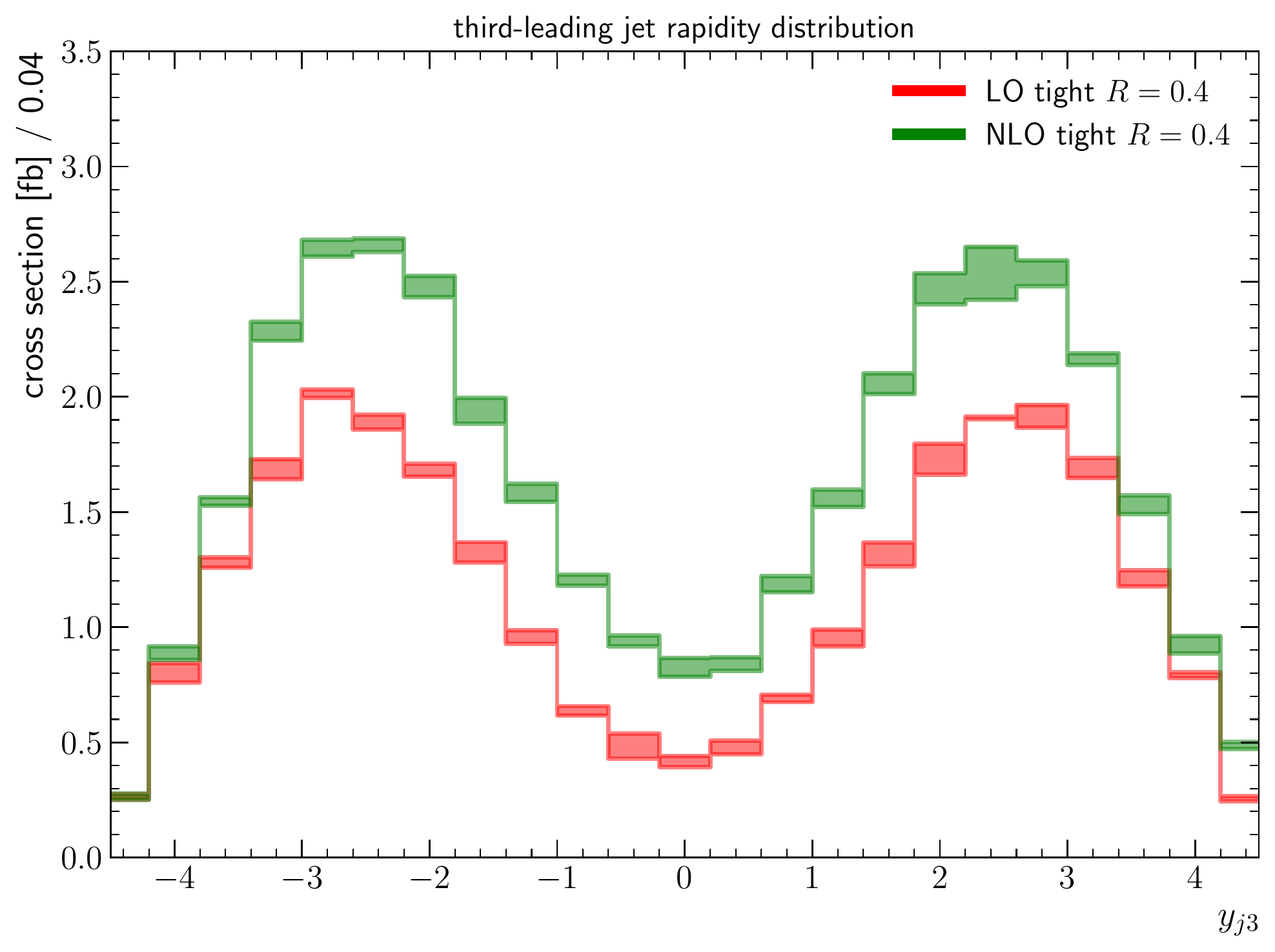}
  \includegraphics[width=0.48\textwidth]{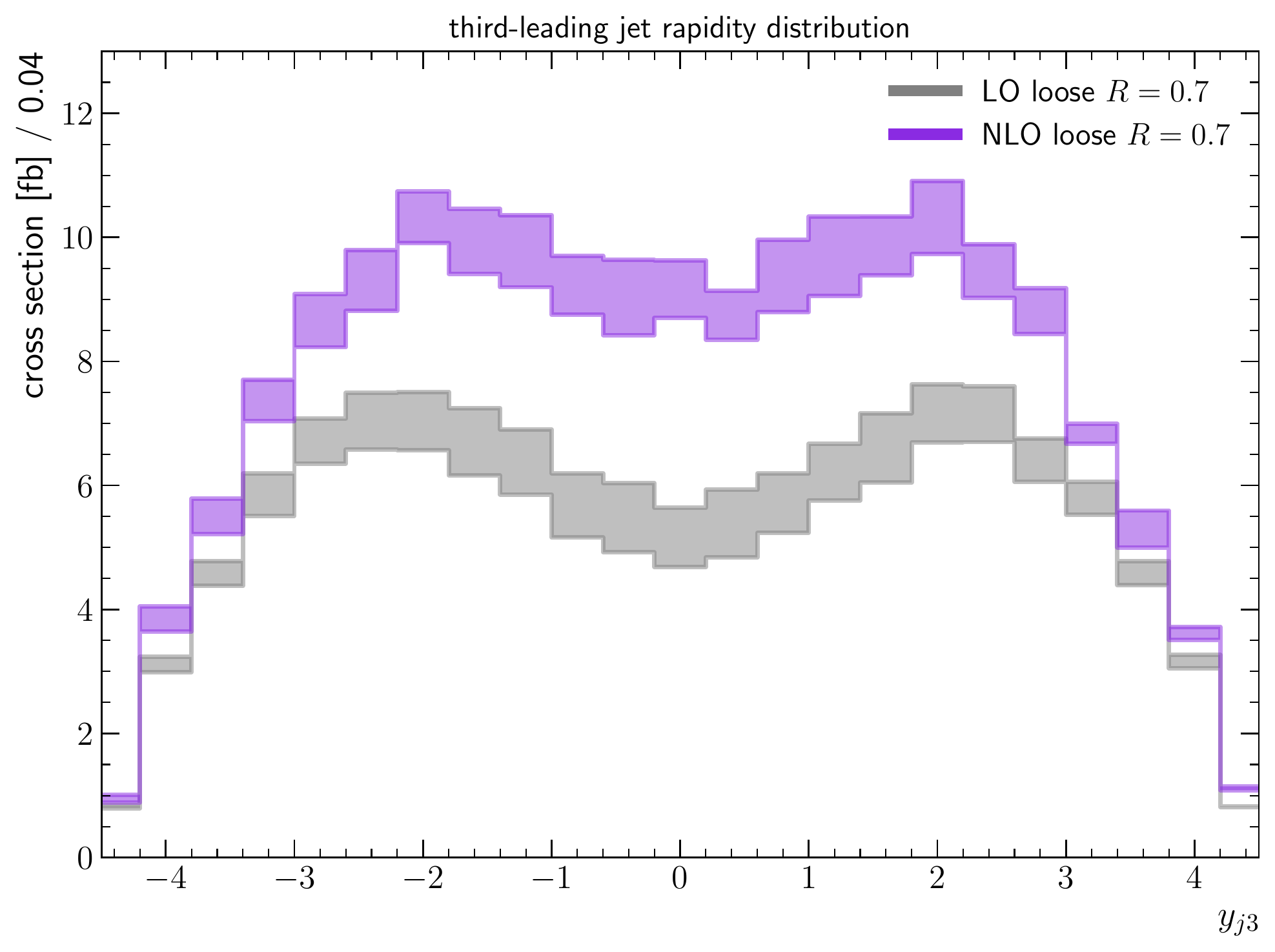}\hfill
  \includegraphics[width=0.48\textwidth]{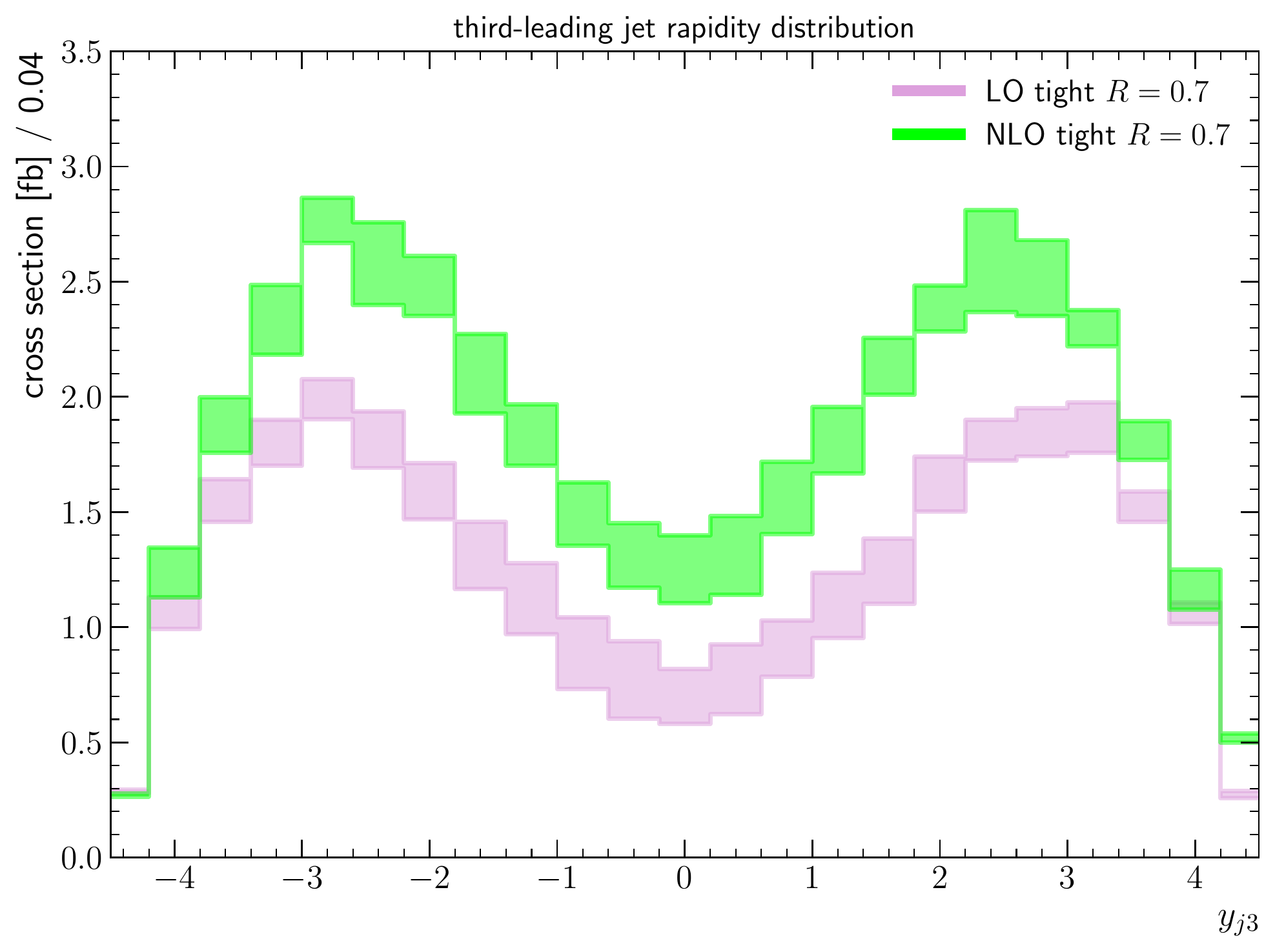}
  \caption{LO versus NLO VBS third-jet rapidity distributions for the loose (left plots) and tight
    (right plots) selection cuts for a jet radius of $R=0.4$ (top
    plots) and $R=0.7$ (bottom plots). The bands constitute the
outermost variations of the MPI model.} 
\label{fig:LOvsNLO}
\end{figure*}

Figure~\ref{fig:LOvsNLO} shows a comparison of the LO and NLO
third-jet rapidity distributions for various setups: loose and
tight selection cuts as described by eqs.~\ref{eq:jetcuts}--\ref{eq:tightcuts}, with jet radius either $R=0.4$ or $R=0.7$. The
bands shown constitute the outermost variations of the MPI model. The
size of this variation is evidently driven by the jet radius, but the
size of the NLO correction is essentially independent of the
setup. Therefore, in what follows, we only examine distributions at NLO.

\subsection{Loose versus tight selection for NLO}

\begin{figure*}[ht]
\centering
\includegraphics[width=0.48\textwidth]{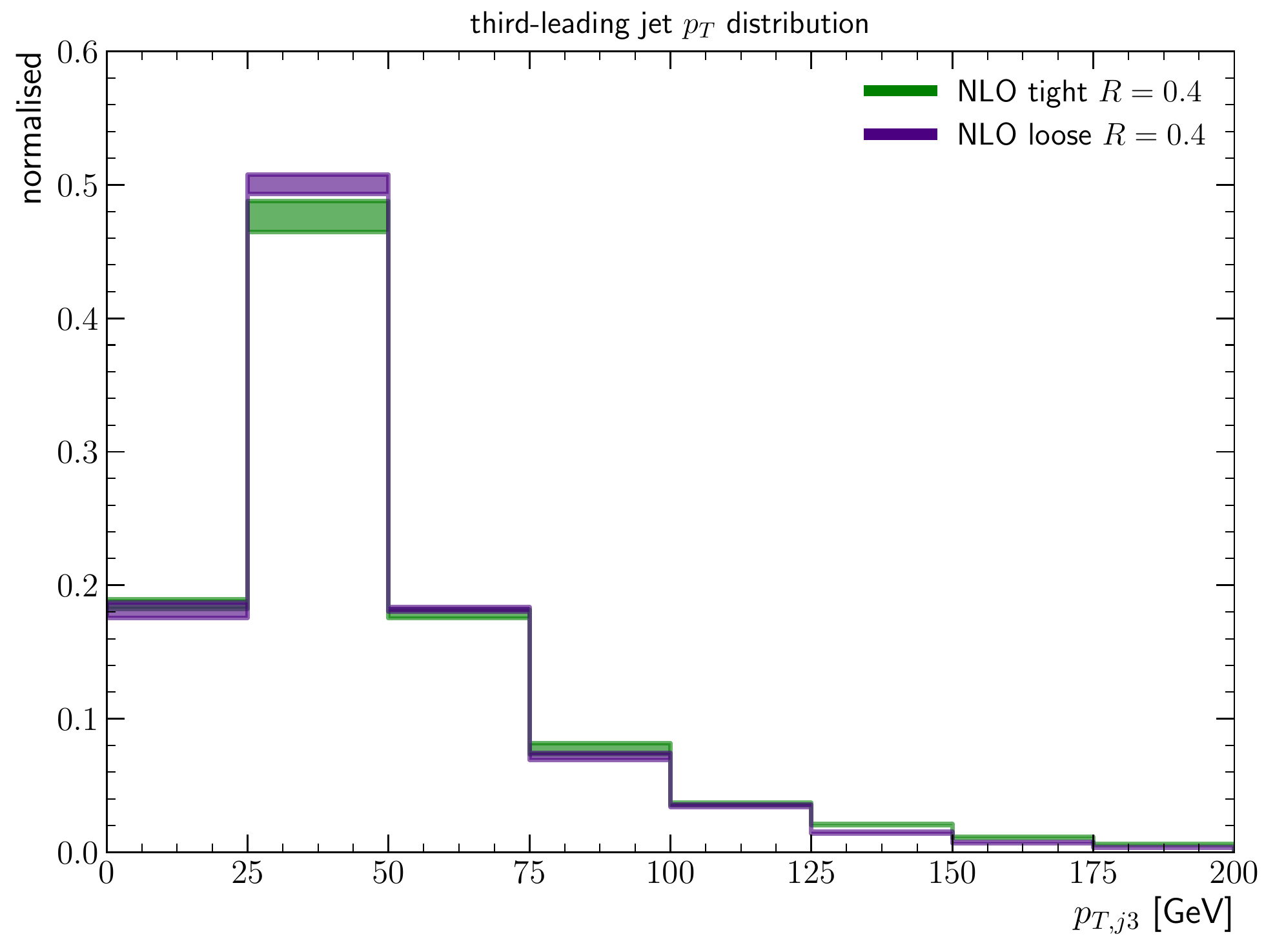}\hfill
\includegraphics[width=0.48\textwidth]{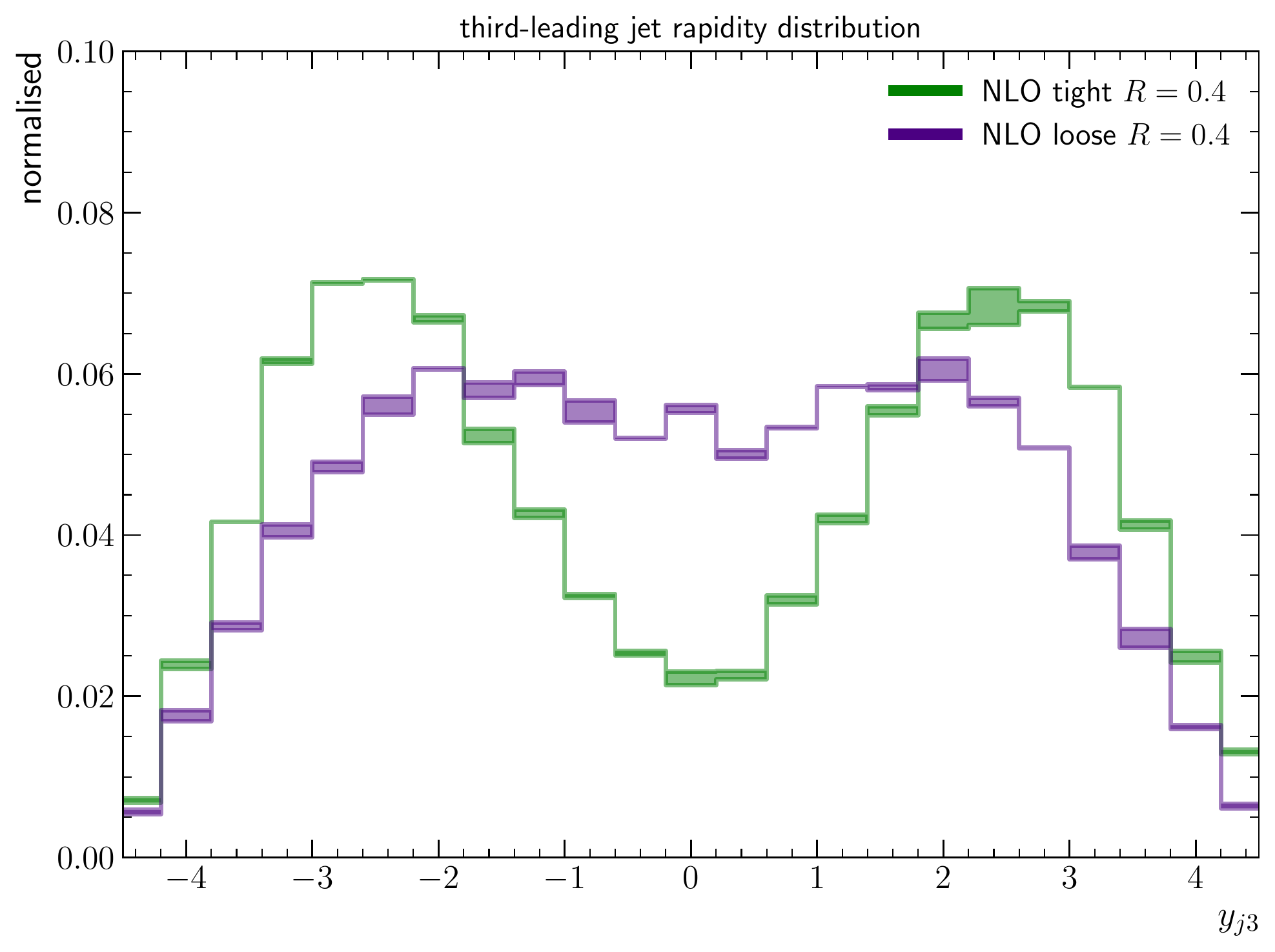}
\includegraphics[width=0.48\textwidth]{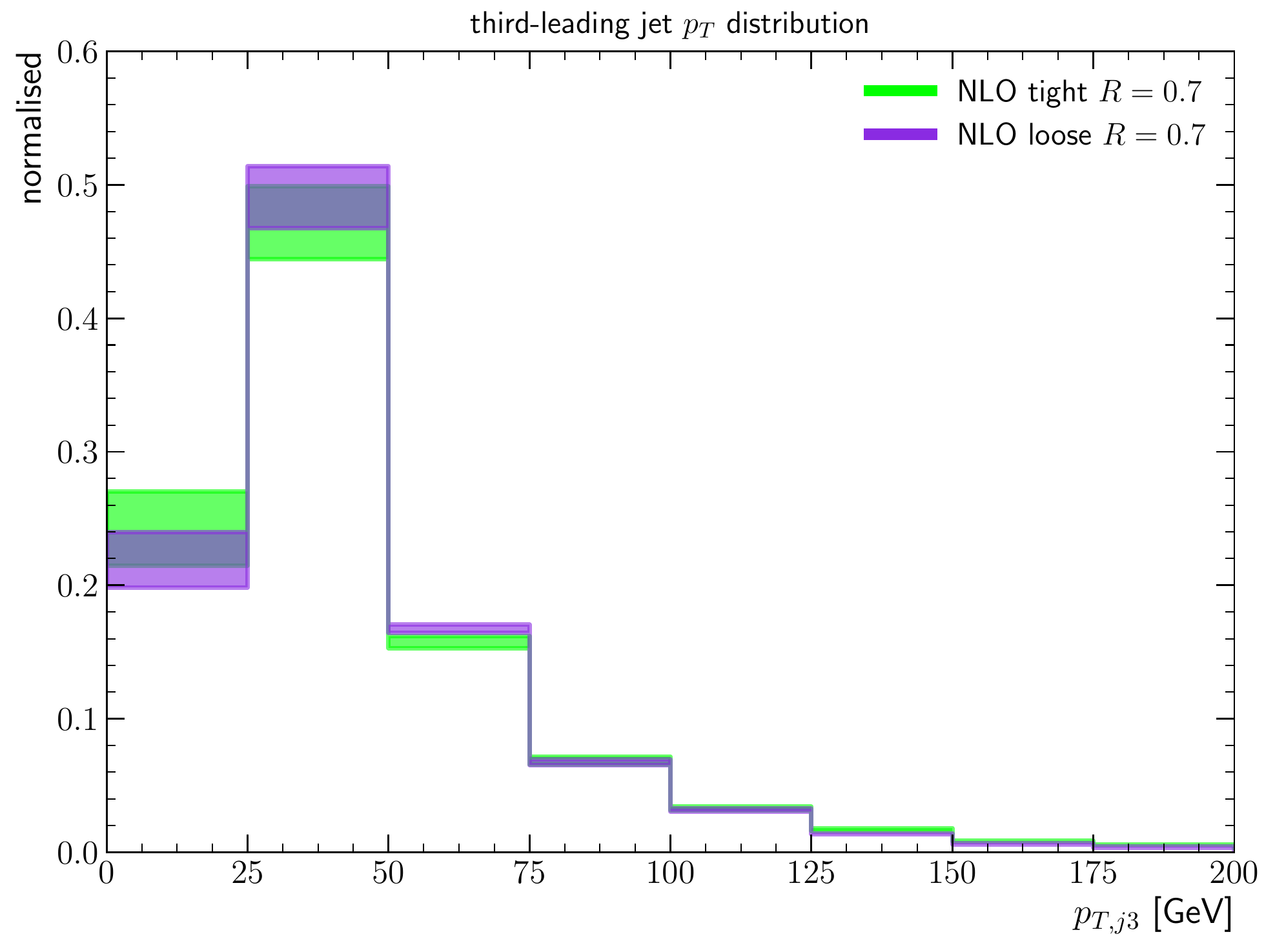}\hfill
\includegraphics[width=0.48\textwidth]{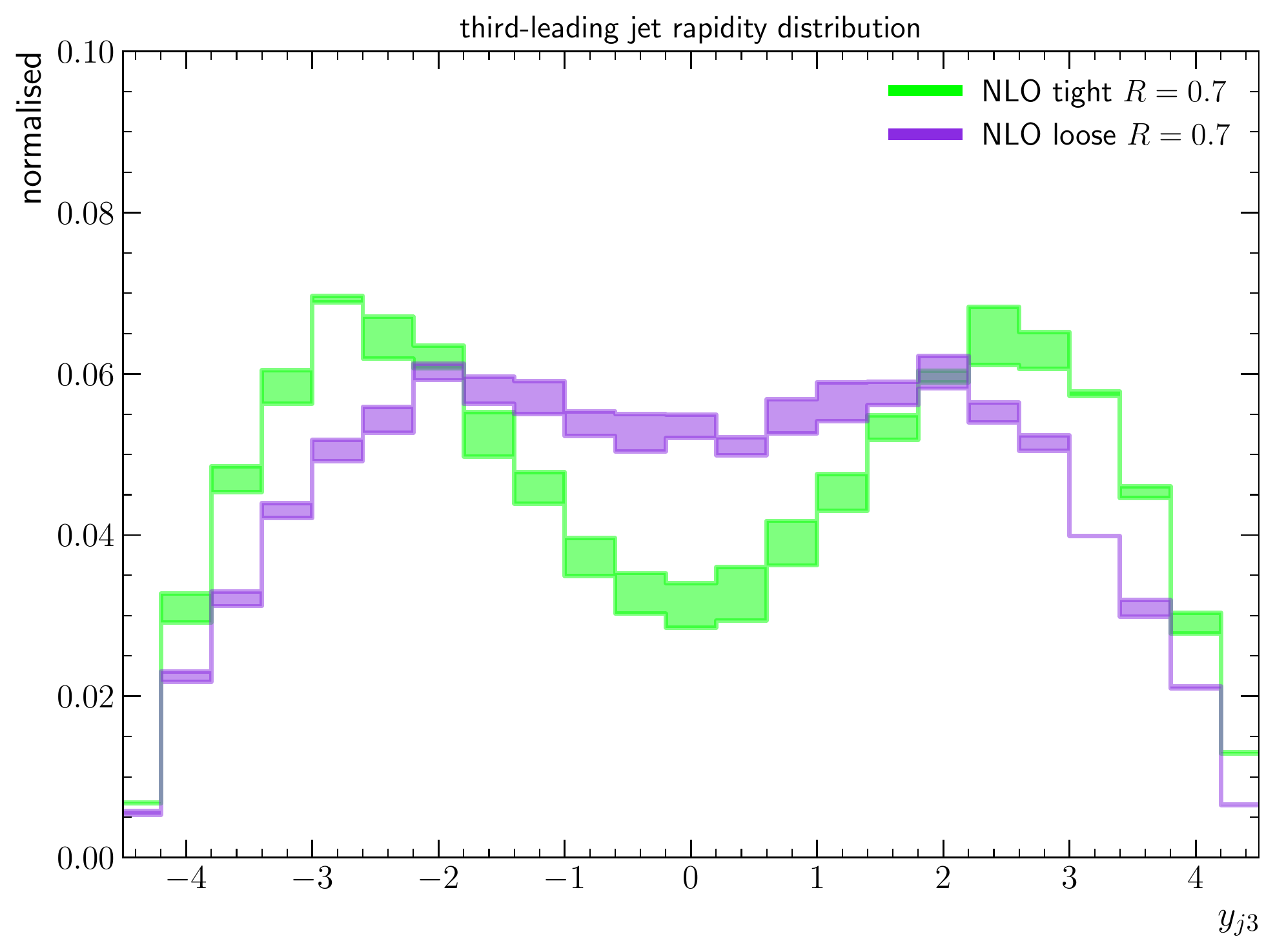}
   \caption{A comparison of NLO VBS observables between the loose and tight selection cuts for a jet
      radius of $R=0.4$ (top plots) and $R=0.7$ (bottom plots). The
      transverse momentum (left plots) and rapidity (right plots) of the third-leading
      jet are shown. The bands constitute the
outermost variations of the MPI model. The histograms have been
      normalised to the same area in each plot.}
\label{fig:NLOTightLoose}
\end{figure*}

In fig.~\ref{fig:NLOTightLoose} we present a comparison of the NLO VBS
third-jet transverse momentum (left) and rapidity (right), between the
loose and tight selection cuts for jet radii of $R=0.4$ and $R=0.7$,
shown in the top and bottom plots, respectively. As before, the bands represent
the outermost variations of the MPI model. It can be seen that a tight
selection cut setup does not reduce the effects of the MPI. On the
contrary, tighter cuts lead to an increase of the uncertainty due to
the MPI variations. This is due to the fact that when applying tight
cuts, the third jet is forced to become more forward than in the case
of loose cuts, effectively making it more sensitive to the soft
activity in the event.

\subsection{MPI versus Shower variations}

\begin{figure*}[ht]
  \centering
  \includegraphics[width=0.48\textwidth]{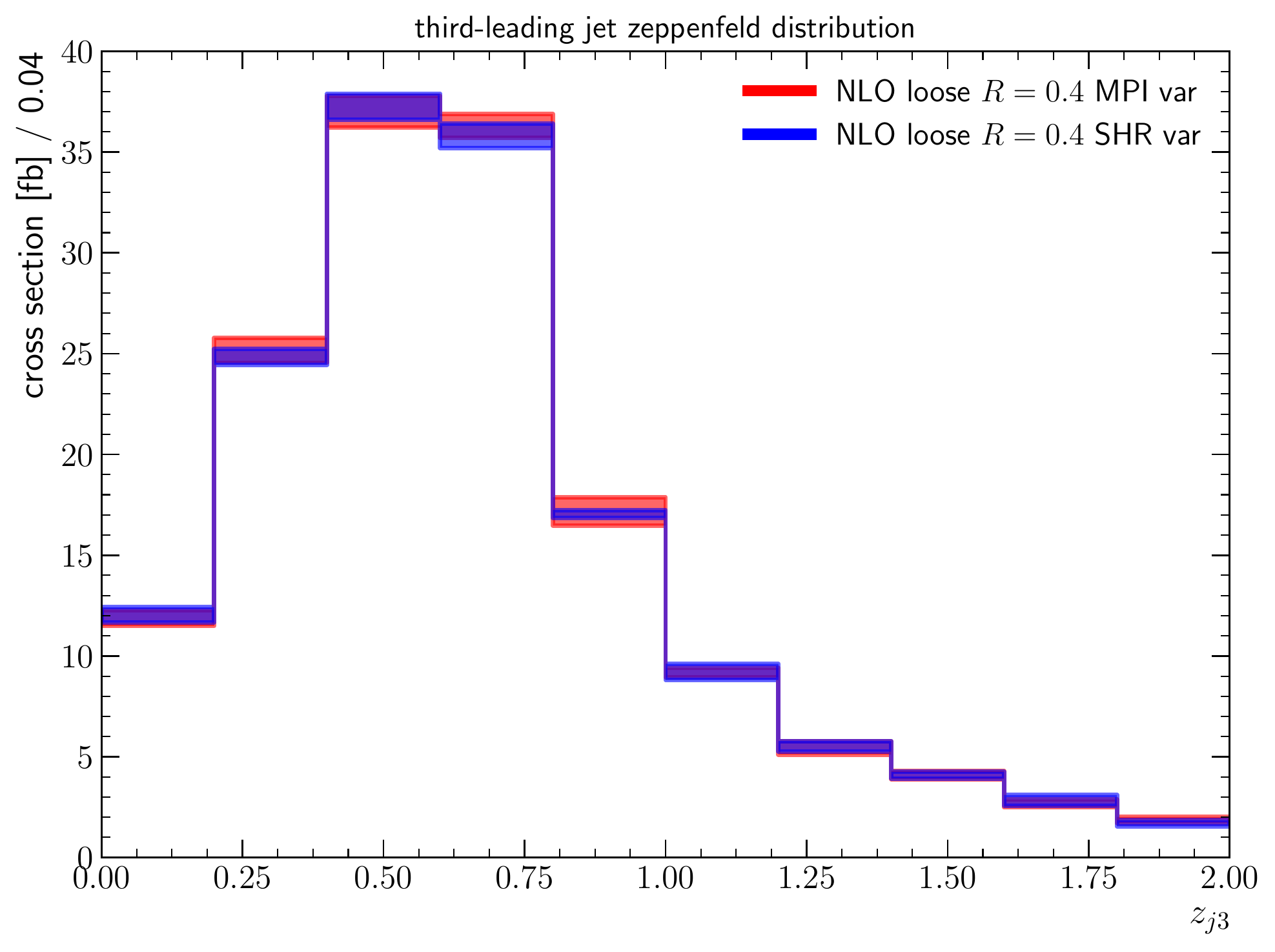}\hfill
  \includegraphics[width=0.48\textwidth]{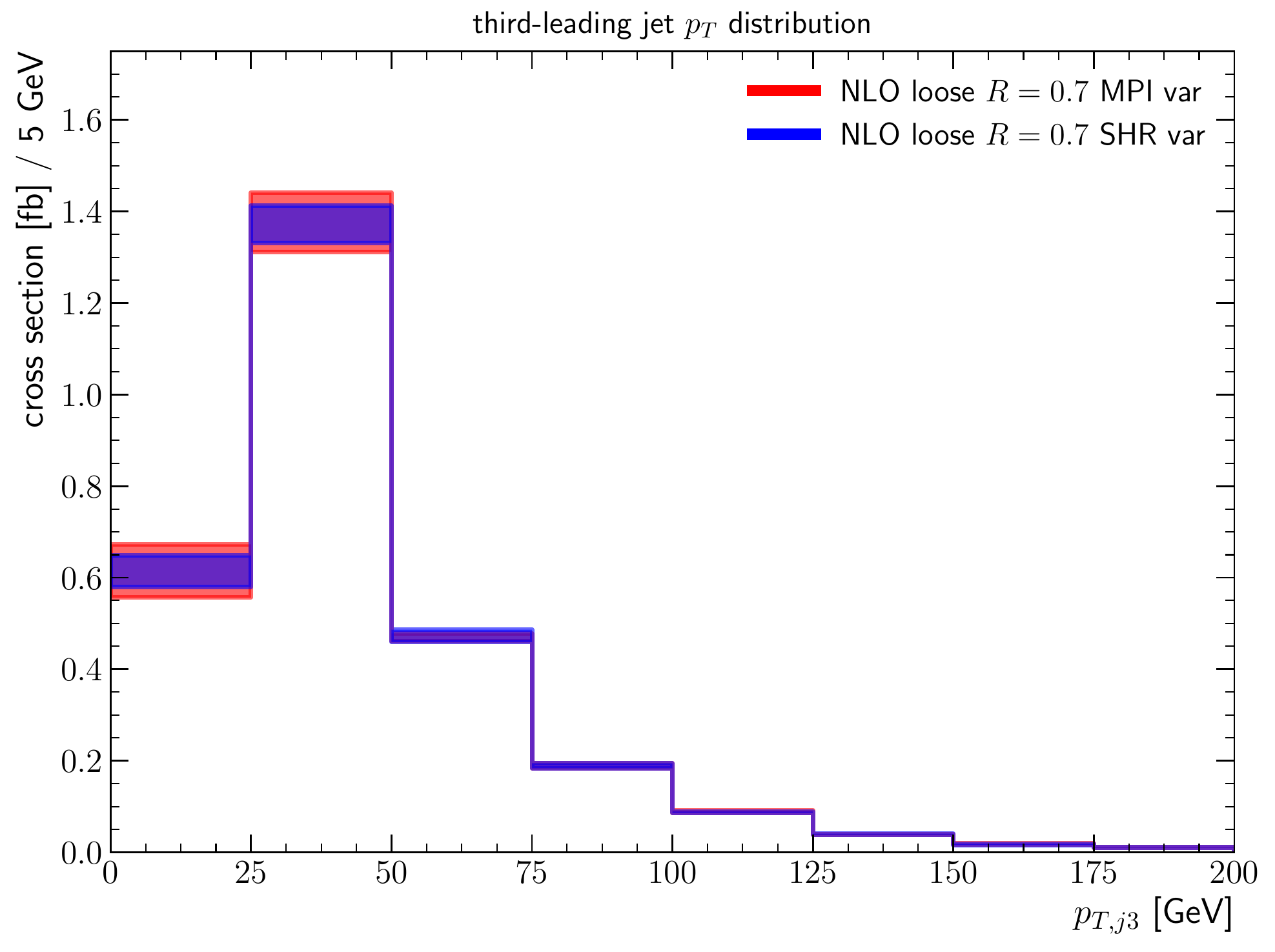}
  \includegraphics[width=0.48\textwidth]{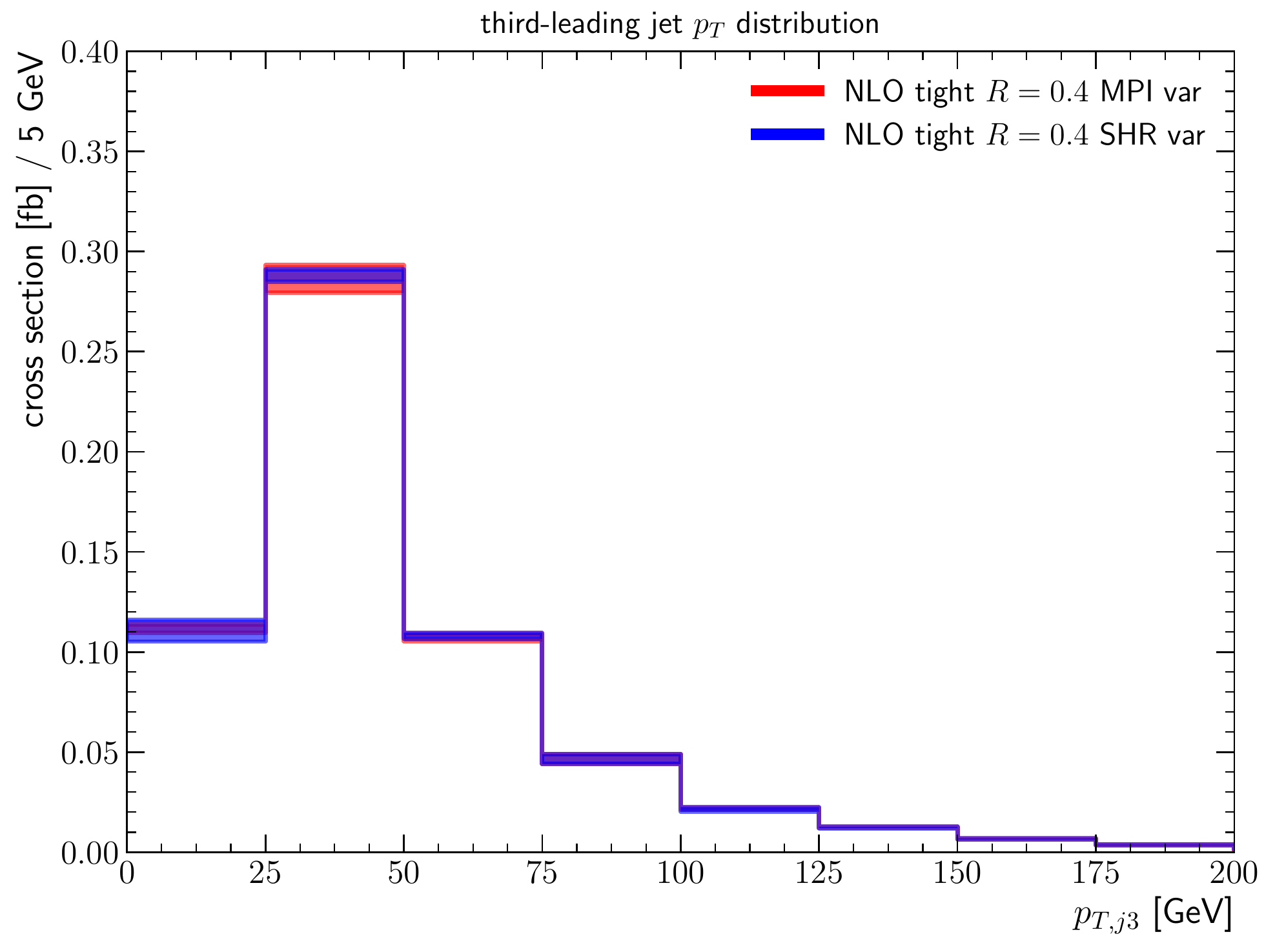}\hfill
  \includegraphics[width=0.48\textwidth]{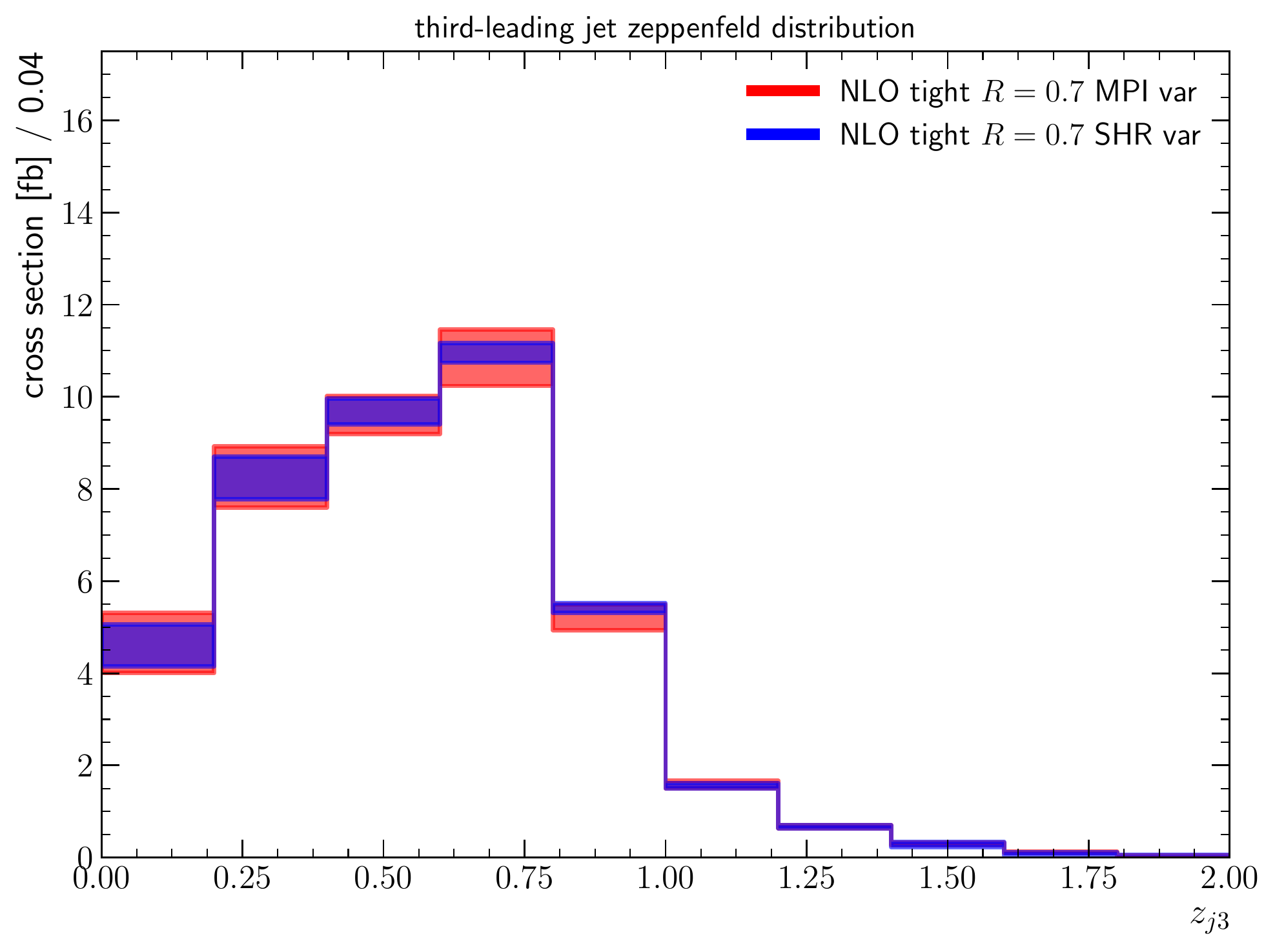}
    \caption{A selection of comparisons of the MPI variations and shower variations
      for NLO VBS observables for the loose (top plots) and tight (bottom
      plots) selection cuts, for a jet radius of $R=0.4$ (left plots)
      and $R=0.7$ (right plots). The transverse momentum and
     Zeppenfeld variable of the third-leading
    jet are shown.}
\label{fig:VarCompNLO}
\end{figure*}

We show in fig.~\ref{fig:VarCompNLO}, a selection of comparisons of
the MPI variations and shower variations for NLO VBS observables for
the loose (top plots) and tight (bottom plots) selection cuts, for a
jet radius of $R=0.4$ (left plots) and $R=0.7$ (right plots). To
reduce the number of plots shown, the transverse momentum and
Zeppenfeld variable of the third-leading jet are shown for specific
setups. One can observe that the MPI variation becomes more important
than the shower variation, particularly for larger jet radii and for
tighter selection cuts.

\section{Conclusion and Outlook}
\label{sec:Conclusion}

In this study we have presented a first assessment of uncertainties
introduced by non-perturbative models in VBF/VBS processes. While
there is no a priori recipe on how to quantify such an
uncertainty from first principles, we have been taking a pragmatic
approach of checking how common sense variations of  $\sim 10\%$
around data typically used in tunes of the MPI and colour reconnection
models extrapolate into a different phase space and class of
observables used in VBF/VBS analysis. We find clear evidence that
these variations become comparable to, and even outrange, those from
the perturbative components in an NLO+PS matched setup. This, in
particular, applies to observables of the third-jet activity which is
used for central jet vetoes and other analyses targeting the colour
structure of the VBF/VBS process. We also find that the variation of
the colour reconnection model, which is of crucial importance to the
description of MPI activity, does not introduce the most significant
variations, and that core parameters of the MPI model are a more
relevant source of uncertainty in this context. Our work demonstrates
that simply evaluating the contribution of a specific non-perturbative
component to the net result of an event generator simulation is not an
accurate analysis of uncertainty and a highly phase-space dependent
phenomenon. A more detailed analysis would also need to account for the
effect of tuning and investigate if there is a tension in between tuning to
classes of observables that do or do not involve the VBF phase
space. Pending availability of relevant data, we leave this exercise
to future work; however one could consider to also do this on the
level of some generated data by re-tuning the model parameters at a
variation of a single parameter and for different classes of
observables.

\section*{Acknowledgments}

This work has been initiated within a Short-term Scientific Mission
funded by the COST action CA16108 ``VBSCAN'' at the University of
Vienna. CB, PK and ST would like to thank the University of Vienna for
their hospitality during this period. PK and SP want to thank the
Erwin Schr\"odinger Institute Vienna for support while this work has
been finalized. SP wants to thank the University of Dresden for
hospitality where further parts of this study have been discussed.  CB
and ST have been funded by the German BMBF project 05H2018 "ErUM-FSP
T02".  This work has also received funding from the European Union’s
Horizon 2020 research and innovation programme as part of the Marie
Skłodowska-Curie Innovative Training Network MCnetITN3 (grant
agreement no. 722104), and in part by the COST actions CA16201
``PARTICLEFACE''. This work was funded in part by the Knut and Alice
Wallenberg foundation, contract number 2017.0036.  We are grateful to
Stefan Gieseke, Frank Siegert and Dieter Zeppenfeld for useful
discussions.

\bibliography{bib}

%% end contents %%%%%%%%%%%%%%%%%%%%%%%%%%%%%%%%%%%%%%%%%%%%%%%%%%%%%%%%%%%%%%%%
%%%%%%%%%%%%%%%%%%%%%%%%%%%%%%%%%%%%%%%%%%%%%%%%%%%%%%%%%%%%%%%%%%%%%%%%%%%%%%%%
\end{document}